\documentclass[a4paper,twoside]{article}

\usepackage{epsfig}
\usepackage{subcaption}
\usepackage{calc}
\usepackage{amssymb}
\usepackage{amstext}
\usepackage{amsmath}
\usepackage{amsthm}
\usepackage{multicol}
\usepackage{pslatex}
\usepackage{apalike}
\usepackage[bottom]{footmisc}
\usepackage[hyphens]{url}
\usepackage{hyperref}
\usepackage{todonotes}
\usepackage{subcaption}
\usepackage{xcolor}
\colorlet{punct}{red!60!black}
\definecolor{delim}{RGB}{20,105,176}
\colorlet{numb}{magenta!60!black}
\usepackage{listings}             
\lstdefinelanguage{json}{
    basicstyle=\footnotesize\ttfamily,
    stepnumber=1,
    numbersep=8pt,
    showstringspaces=false,
    breaklines=true,
    frame=lines,
    literate=
     *{0}{{{\color{numb}0}}}{1}
      {1}{{{\color{numb}1}}}{1}
      {2}{{{\color{numb}2}}}{1}
      {3}{{{\color{numb}3}}}{1}
      {4}{{{\color{numb}4}}}{1}
      {5}{{{\color{numb}5}}}{1}
      {6}{{{\color{numb}6}}}{1}
      {7}{{{\color{numb}7}}}{1}
      {8}{{{\color{numb}8}}}{1}
      {9}{{{\color{numb}9}}}{1}
      {:}{{{\color{punct}{:}}}}{1}
      {,}{{{\color{punct}{,}}}}{1}
      {\{}{{{\color{delim}{\{}}}}{1}
      {\}}{{{\color{delim}{\}}}}}{1}
      {[}{{{\color{delim}{[}}}}{1}
      {]}{{{\color{delim}{]}}}}{1},
}
\usepackage{SCITEPRESS}     

\begin{document}

\title{How to Design a Blue Team Scenario for Beginners on the Example of Brute-Force Attacks on Authentications}

\author{\authorname{Andreas Eipper\sup{1} and Daniela Pöhn\sup{1}}
\affiliation{\sup{1}Universität der Bundeswehr München, Neubiberg, Germany}
\email{\{firstname.familyname\}@unibw.de}
}


\abstract{Cyber attacks are ubiquitous and a constantly growing threat in the age of digitization. In order to protect important data, developers and system administrators must be trained and made aware of possible threats. Practical training can be used for students alike to introduce them to the topic. A constant threat to websites that require user authentication is so-called brute-force attacks, which attempt to crack a password by systematically trying every possible combination. As this is a typical threat, but comparably easy to detect, it is ideal for beginners. Therefore, three open-source blue team scenarios are designed and systematically described. They are contiguous to maximize the learning effect.}

\onecolumn \maketitle \normalsize \setcounter{footnote}{0} \vfill

\section{\uppercase{Introduction}}
\label{sec:introduction}

The world is becoming more and more connected through digital systems. According to \cite{Nate}, users have between 80 and 150 online accounts, including social media, online banking, and many more. With the help of publicly available information and basic hacking skills, many authentication systems can be infiltrated or compromised, giving attackers access to users' personal information. Each user should protect their accounts with individual secure passwords. In practice, users tend to use single passwords for multiple services. This increases the impact of a security incident. Simple passwords, such as \texttt{123456}, \texttt{test1}, \texttt{qwerty}, \texttt{iloveyou} and others in wordlists like \texttt{rockyou.txt}, can be easily cracked. Social engineering and online research, also called open source intelligence (OSINT), help hackers figure out valid passwords. The information obtained can automate brute-forcing as the attack progresses. As a result of various incidents and attacks (e.\,g., data leaks, brute-force attacks, and phishing), lists of user names and passwords can be found on the Internet. Credential stuffing is, for example, an attack that automatically tries these stolen credentials for other services.

The aim of this paper is to impart knowledge and skills in the field of detection and countermeasures to participants in cyber range training courses by designing and implementing open-source blue team scenarios with the topic of brute-force attacks on authentication in a web application. Three subsequent scenarios each cover a specific part of the learning content and lay the foundation for beginners and those interested in protection against brute-force attacks.

Therefore, this paper contributes 1) the design process for beginner scenarios; 2) a description of the overall training setting and each scenario in a generic way; 3) an evaluation based on a training session.

The paper is structured as follows: Related work is introduced in Section~\ref{sec:relatedwork}. Next, the concept of the three beginner scenarios is outlined. This is the basis for the practical implementation in Section~\ref{sec:implementation}, which is then tested with students, as described in Section~\ref{sec:evaluation}, and discussed in Section~\ref{sec:discussion}. Last but not least, the paper is concluded and future work is given in Section~\ref{sec:conclusion}.

\section{\uppercase{Related Work}}
\label{sec:relatedwork}

This chapter evaluates related work to brute-force detection and prevention as well as cyber training and training scenarios. To the best of our knowledge, there is no training for learning about brute-force detection and prevention.

\paragraph{Brute-Force Detection and Prevention:} Applying detection techniques is important to detect anomalous behavior early and minimize its impact on the network. One such application is Wireshark, as described by \cite{bruteforceanawithws}. The authors reason that the automatic ban function of FileZilla is not enough to stop a brute-force attack. Therefore, they recommend deactivating the targeted account. This is possible if only single or selected accounts are under attack. Other important sources for detection are log files. Apache hypertext transfer protocol (HTTP) server offers different log files, including \texttt{error.log} (diagnosis information and errors) and \texttt{access.log} (processed requests). Based on these log files, different attacks can be noticed by personnel if the attacks are known \cite{logfilelitrech,apacheaccerr}. However, it also requires training to detect attacks, either manually or with the aid of tools. \cite{LopezAraiza2017} describe tools for network security and forensics including Fail2Ban, which provides detection and ban functionalities.

\paragraph{Cyber Training:} To achieve pedagogical added value, it is important to design the training competently. \cite{gamifcation} provide guidelines on how training can be effectively designed using the principle of serious games. The training is intended to reproduce realistic environments that require strategic and adversarial thinking. \cite{CyberRangeDocUniBw} categorize cyber training. \cite{CyberRangeDefUniBw} examined several training courses to determine their added value for network defense. The authors conclude that the representation of realistic and comprehensible attack scenarios with different patterns leads to a high learning effect. The training should be accompanied by specialist staff. A high proportion of practical exercises can reduce the duration of training while achieving higher learning goals. In addition, different online learning platforms provide gamified real-world labs as evaluated by \cite{steiner}.

\paragraph{Training Scenarios:} \cite{AllgemeineAnalyse} specify cyber attacks by applying the diamond model, consisting of an adversary, capability, infrastructure, and victim. These categories can be applied for training scenarios as well, but are not specific enough. \cite{6392562} give an overview of game design for cyber security training, whereas \cite{9527946} analyze evaluation metrics for cyber security training. \cite{8166396} propose amongst other things a workflow diagram for training exercises, describing the main steps objectives; environment configuration; design and deploy; train test score; and analyze evaluate adjust. Although these steps can be applied, they are rather generic. The design and use cases of the KYPO cyber ranges are presented by \cite{icsoft17}. In their documentation, \cite{kypo} state how to use their patterns to design use cases and workflows. Even though it is a systematic approach, it is focused on their environment. Similarly, \cite{9454094} propose a domain-specific language based on the MITRE ATT\&CK framework for dynamic training in cyber range environments. The authors describe the classification, environment, execution, and evaluation in a systematic, but also high-level way.

\paragraph{Summary:} Skills in network analysis is a core task for blue teams. Training designed for beginners is hardly described. We did not find any training on brute-force attacks related to authentication. Therefore, we provide a step-by-step guide and a generic description of the scenarios, which can be repurposed for other training settings.

\section{\uppercase{Concept}}
\label{sec:concept}

This chapter outlines the theoretical composition and structure of the designed beginner scenarios. The description of the structure of the three subsequent scenarios is based on \cite{CyberRangeDocUniBw}. Thereby, the audience, training environment, training setup, and technical setup are defined in general. In addition, the scenario, goals, and resources are explained in more detail. This is supplemented with a possible procedure based on related work, described in Section~\ref{sec:relatedwork}. Based on the received feedback, see Section~\ref{sec:evaluation}, prerequisites are necessary to pre-access.

\begin{figure*}[!htpb]
\begin{subfigure}[c]{0.3\textwidth}
    \includegraphics[width=0.98\linewidth]{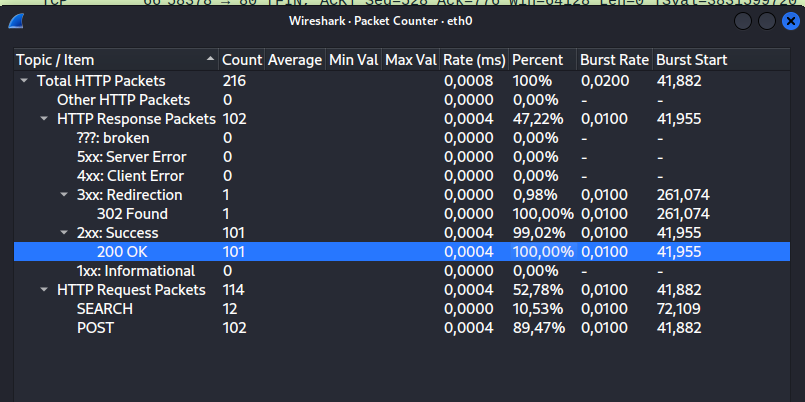}
    \subcaption{Scenario 1: Package analysis in Wireshark}
    \label{fig:wireshark}
\end{subfigure}
\begin{subfigure}[c]{0.38\textwidth}
    \includegraphics[width=0.98\linewidth]{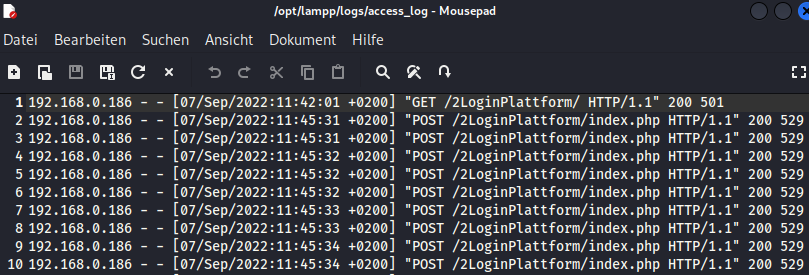}
    \subcaption{Scenario 2: Excerpt of log files}
    \label{fig:logs}
\end{subfigure}
\begin{subfigure}[c]{0.3\textwidth}
    \includegraphics[width=0.98\linewidth]{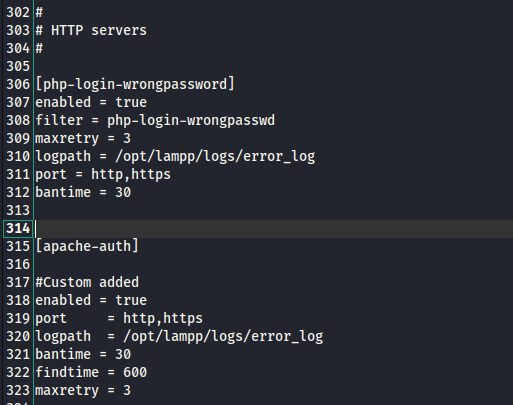}
    \subcaption{Scenario 3: Rules in Fail2Ban}
    \label{fig:fail2ban}
\end{subfigure}
\caption{Scenarios}
\label{fig:scenarios}
\end{figure*}

\subsection{Overview}

\paragraph{Audience:} The target audiences are students and other interested persons with basic cyber security knowledge. Thereby, the prerequisite consists of the knowledge gained in a bachelor's program. Hence, the sector is described as universities and the proficiency level is the related audience. The training has the purpose to increase the ability to detect brute-force attacks on authentication of websites, thereby, using log files and Wireshark, resp. understanding of intrusion prevention.

\paragraph{Training Environment:} The training is carried out in dedicated computer rooms or in our cyber range. With the goal to understand brute-force attacks on authentication, their detection, and prevention, the material is divided into three distinct scenarios. Each scenario covers a certain aspect to learn basic knowledge:

\begin{enumerate}
    \item Wireshark: As a visual instrument for analyzing and graphically processing data logs~\cite{wiresharkbegr,bruteforceanawithws}. In addition, Wireshark could be used in advanced scenarios with no suitable log files.
    \item Log Files: As a system-related and fast source of information and supplement to Wireshark. Log files are utilized in more advanced tools. \cite{logfilelitrech,apacheaccerr}
    \item Fail2Ban: As a simple tool for intrusion prevention~\cite{LopezAraiza2017}.
\end{enumerate} 

The first two scenarios cover intrusion detection. In the last scenario, Fail2Ban is presented as an intrusion prevention system (IPS) for blue teams. Thereby, the participants first see the malicious traffic graphically, before analyzing the in-going traffic with the command line. Last but not least, a prevention possibility is explained. The topics are coherent and can, therefore, be dealt with in one go for maximum learning effect. For added pedagogical value, the diamond framework~\cite{AllgemeineAnalyse} and the cyber kill chain can be used as an aid. In addition, follow-on topics can be discussed.

\paragraph{Training Setup:} No scoring is set up in these three basic scenarios as the primary goal is to introduce the participants to the concepts and handling of Kali and its tools. Scoring can be applied in advanced scenarios to challenge the participants. The participants act as a blue team in single mode at a general level.

\paragraph{Technical Setup:} The scenarios try to rebuild a simplified realistic environment structure. The deployment is on-premise with a resp. several virtual machines (VMs) due to the architecture of our TAME range, as outlined by~\cite{8992693}. Although VMware is the primary installation, VirtualBox or other products could be imported.

\paragraph{Description:} The training concept can be summarized as in Listing~\ref{lst:training}. First, a description of the scenarios (name, goal, scoring, and environment) is given (Lines 3-7). Then the set of scenarios with each scenario, intended tool, and goal are generally stated (Lines 8-20). This description language is based on related work in Section~\ref{sec:relatedwork}. It tries to balance detailed information and generic representation. The description utilizes JavaScript Object Notation (JSON) as a prominent candidate for descriptions on the Internet, similar to the related work. Depending on the cyber range, either automated provisioning with tools such as Vagrant and Puppet or VMs are set up \cite{8992693}. In consequence, the descriptions either result (with translation) in automated provisioning or can be used to build the VMs. The scenarios and the technical setup need to be detailed in further descriptions.

\begin{lstlisting}[language=json,caption={Training description},label=lst:training]
{
 "training": {
  "description": {
   "name": "Brute-Force AuthN",
   "goal": "analyze",
   "scoring": "none",
   "environment": "cyber range",
   "scenarios": [
    { 
     "scenario": "network traffic",
     "tool": "Wireshark",
     "goal": "understand",
    }, {
     "scenario": "logging",
     "tool": "log files",
     "goal": "apply",
    }, {
     "scenario": "IPS",
     "tool": "Fail2Ban",
     "goal": "analyze",
    }
   ]
  },
 }
}
\end{lstlisting}

\subsection{Scenario 1: Wireshark}

\paragraph{Training Environment:} The scenario provides the participants with a VM to work on. The aim is to discover abnormalities in the packet capture (PCAP) files. The following learning objectives are aimed for in this scenario.
\begin{itemize}
    \item Wireshark basics:
    \begin{itemize}
        \item Operation of the tool.
        \item Filter application.
        \item Statistics creation and evaluation.
    \end{itemize}
    \item Detection of simple brute-force attacks.
    \item Awareness of log files.
\end{itemize}

\paragraph{Training Setup:} The lecturer discusses the basics of Wireshark and brute-force attacks, in order to create a theoretical understanding. The participants search for and open the PCAP file in Wireshark and try to recognize the attack, shown in Figure~\ref{fig:wireshark}. Then, mitigation strategies are discussed.


\paragraph{Technical Setup:} The participants use a stored PCAP file.

\subsection{Scenario 2: Log Files}

\paragraph{Training Environment:} The aim is to discover anomalies in the log files and to include them in the analysis results of scenario 1. The following learning objectives are aimed for in this scenario.
\begin{itemize}
    \item Basics Log Files:
    \begin{itemize}
        \item Types of log files.
        \item Locations of log files on Debian distributions.
        \item Evaluation of log files.
    \end{itemize}
    \item Detection of brute-force attacks.
    \item Awareness of log files.
\end{itemize}

\paragraph{Training Setup:} After an introduction to log files, the participants search for and open the log files and analyze them accordingly (pattern, IP addresses, in combination with the PCAP file).


\paragraph{Technical Setup:} The participants use the according log files \texttt{access.log} and \texttt{error.log}.

\subsection{Scenario 3: Fail2Ban}

\paragraph{Training Environment:} The goal is to activate the Fail2Ban configuration and demonstrate the capabilities of the tool in a practical demonstration. This scenario has the following learning objectives.
\begin{itemize}
    \item Intrusion Prevention Basics:
    \begin{itemize}
        \item Functionality.
        \item Areas of application.
    \end{itemize}
    \item Basics Fail2Ban:
    \begin{itemize}
        \item Theoretical functionality.
        \item Practical implementation.
    \end{itemize}
\end{itemize}

\paragraph{Training Setup:} The lecturer summarizes the results from the previous scenarios to create a basis for discussions on ways to combat brute-force attacks. After collecting and evaluating ideas and suggestions, Fail2Ban and intrusion prevention, in general, are presented. Once the theoretical basis has been created, Fail2Ban is installed and configured. The properties of the \texttt{local.jail} file and the necessary commands for activation are shown here, see Figure~\ref{fig:fail2ban}. The functionality is demonstrated using an example attack. Finally, possible errors and weaknesses of Fail2Ban are discussed with the participants.


\paragraph{Technical Setup:} The participants use Fail2Ban with its configuration and log files.

\section{\uppercase{Implementation}}
\label{sec:implementation}

The implementations and technical precautions required to carry out the training are described in this section. The implementation can be summarized as in Listing~\ref{lst:implementation}. First, a description of the implementation is given (Lines 3-5). Then the environment featuring the different teams, i.\,e., blue team (Lines 7-19), red team (Lines 20-23), and yellow team (Lines 24-27), is summarized. This especially includes platforms, tools, IP addresses, and resources.

\begin{lstlisting}[language=json,caption={Implementation description},label=lst:implementation]
{
 "training": {
  "description": {
   "name": "Brute-Force AuthN",
    "scenarios": "3",
  }, "environment": {
   "blueteam": {
    "platform": "Kali Linux",
    "tools": ["Wireshark", "log files", "Fail2Ban"],
    "ip": "192.168.1.10",
    "infrastructure": {
     "name": "WebApp",
     "goal": "web application",
     "tools": ["PHP", "Apache", "phpMyAdmin"],
     "sources": ["PHP pages", 
        "/var/log/*", 
        "/opt/lampp/logs/*",
        "/etc/fail2ban/jail.local"],
    },
   }, "redteam": {
    "platform": "Kali Linux",
    "tools": ["Firefox", "Burp Suite", "FoxyProxy"],
    "ip": "192.168.2.1-100",
   }, "yellowteam":{
    "platform": "Windows 11",
    "tools": ["Chrome", "Selenium"],
    "ip": "192.168.2.1-100",
   },
  }
 }
}
\end{lstlisting}

As a basis for the blue and red team platforms, VMs are created with Oracle VM VirtualBox and Kali Linux as the operating system. Alternatively, Debian could be used for blue teams.

\begin{figure*}[!htpb]
\begin{subfigure}[c]{0.4\textwidth}
    \includegraphics[width=0.98\linewidth]{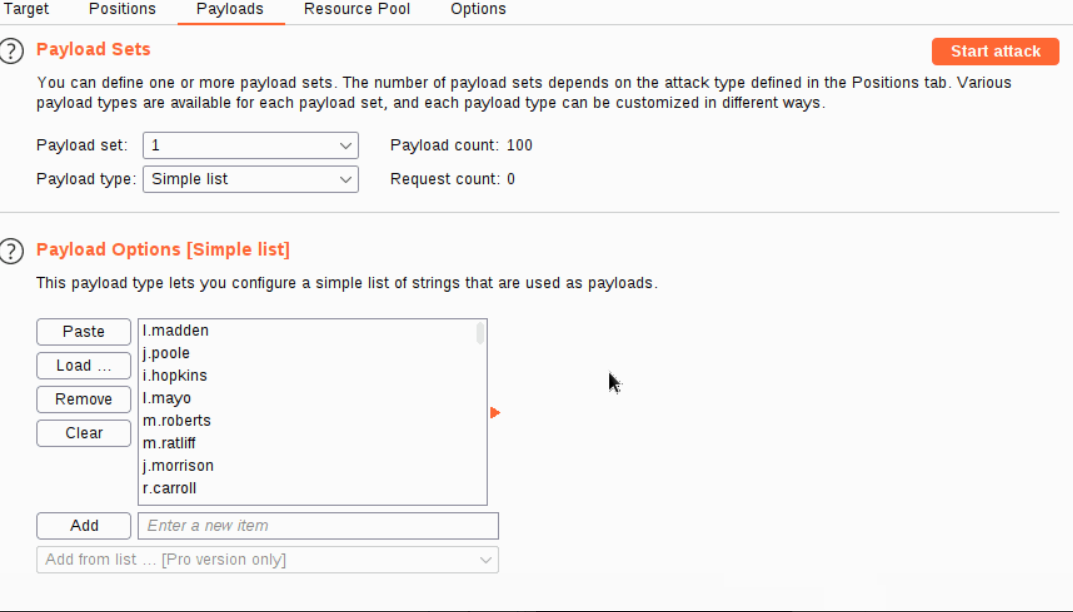}
    \subcaption{Data in Burp Suite Intruder}
    \label{fig:burp}
\end{subfigure}
\begin{subfigure}[c]{0.56\textwidth}
    \includegraphics[width=0.98\linewidth]{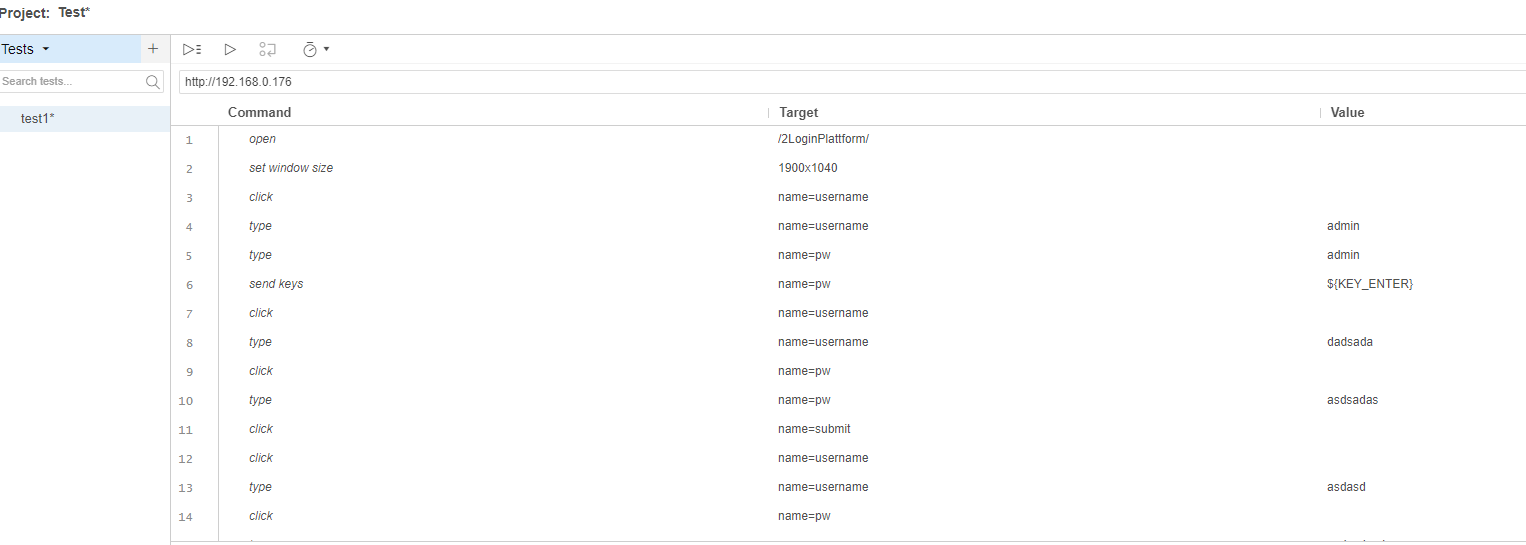}
    \subcaption{Recorded sessions in Selenium}
    \label{fig:selenium}
\end{subfigure}
\caption{Implementation}
\label{fig:implementation}
\end{figure*}

The web application of the infrastructure is created with the LAMPP/XAMPP framework (PHP, Apache, MariaDB, and phpMyAdmin) by \cite{xampp}. It consists of five PHP files: a login page, homepage, registration page, logout page, and a connection page to the database. The data required for the web application is stored in the database. The log files can be found at the default location. In addition, several tools for the analysis are available or can be installed, such as Fail2Ban.

In order to simulate the attacks (red team), the Burp Suite~\cite{BurpSuiteInfos} is utilized. Burp Suite is a tool developed by Portswigger Web Security with the ability to act as a proxy for manual testing of web applications. Other applications such as THC Hydra, Medusa, and Patator as shown by \cite{tools} could be used instead but Burp Suite provides a graphical interface and is often applied for website testing. Burp Suite is triggered by the browser plugin \cite{foxy}. Burp Suite Proxy and Intruder~\cite{thmintruder} catch the request to the target. The attacker then can modify it to start the brute-force attack as shown in Figure~\ref{fig:burp}. With changing IP addresses through the usage of proxies, Tor, or virtual private networks (VPNs), a distributed attack can be imitated, if no distributed setting is available. Alternatively, editing files is possible.

The tool selected to generate ordinary network traffic is Selenium~\cite{seleniuminf}. Originally, Selenium is a portable testing framework primarily focused on testing web-based applications. It provides a record-playback feature that helps in recording test case executions and allows testers to play them back at a later time as shown in Figure~\ref{fig:selenium}. Thereby, simple user behavior can be simulated~\cite{9155736,10.1145/3055186.3055188}. A new test is first started and then a login process is recorded. These and other tests can be run as often as you like at later times, thereby, simulating regular user behavior. This has the advantage of a simple generation. On the other hand, it replays without much variation.

\section{\uppercase{Evaluation based on Training Session}}
\label{sec:evaluation}

We evaluate the training scenarios and their description through a small-scale training session with students. In order to test the learning effects of the scenarios, a training session was conducted. The participants were predominantly male, young, and technical-savvy. They had a rudimentary knowledge of cyber security. The following criteria were evaluated during the training and subsequent assessment:

\begin{description}
    \item[Sources of information:] The participants had material about brute-force attacks. The material was described as comprehensive and intuitive in terms of structure and writing style. During the training, it turned out that the amount of information material becomes confusing for beginners due to the complexity and the missing knowledge about Linux and the command line.
    \item[Training structure:] The structuring was emphasized positively. However, the training revealed slower learning progress than expected. Nevertheless, all participants successfully completed the training within the allocated and some buffer time. One possible option could be to separate the scenarios based on their properties. Another option would be to add an informative session beforehand to explain the theory. This though would extend the training time in total~\cite{CyberRangeDefUniBw}. Last but not least, the Linux OS Kali and the command line could be introduced in another session. Several courses about Kali can be found online, including the official course PEN-103 by \cite{pen}.
    \item[Training process:] The guide was positively received and was effective according to all participants. Here too, however, the beginner-typical deficits came clear. Basics such as operating the OS Kali, in particular using the command line to configure Fail2Ban, also had to be explained.
\end{description}

In summary, the training was positively received by all. The learning effect and awareness of cyber security were evident. Due to the lack of basic understanding, it is advisable to verify the required knowledge and locate time accordingly. The preferred option for our training session is to include a pre-session about Linux and Kali in particular, where the students can familiarize themselves with the command line.

\section{\uppercase{Discussion}}
\label{sec:discussion}

Based on the concept and practical implementation of the subsequent training sessions, we discuss major elements of our beginner training.

\paragraph{Design of Beginner Scenarios:} By applying the cyber taxonomy~\cite{CyberRangeDocUniBw}, we used a systematic approach to design and describe our beginner scenarios. Based on the goals and learning skills, three subsequent scenarios were created. Whereas several cyber range architectures and the usage of cyber training for educating students are published, we found no beginner scenarios in the literature. This workflow can be applied to future scenarios. Although it suited our purpose, a refined version might be required for advanced cyber range scenarios. In addition, prerequisites need to be verified.

\paragraph{Network Traffic:} Realistic synthetic network traffic is an important element of cyber training. \cite{10.1145/3488375} provide a categorization of different tools and methodologies applied. The overview shows that there is no standard solution so far. We applied Selenium, which is not included in the overview, as we focused on web application authentication and the replay of recorded sessions suits the purpose. Nevertheless, it is not realistic enough for more advanced scenarios.

\paragraph{Structured Description Language:} The designed subsequent scenarios target beginners and, therefore, do not consist of a complicated setting such as in \cite{kypo}. Even though the first approaches presented structured description languages, no standard has been established yet. In contrast to the examples accompanying these structured description languages, our scenarios are comparably easy and, therefore, do not fit. We tried to adapt them to our scenarios. Contrary, our description language fits our purpose but is not suited for complex settings. Therefore, further work is required toward an advanced description language for different scenarios.

\paragraph{Content:} According to \cite{10.1145/1414558.1414607}, web application security should be integrated into IT curriculum. Our subsequent scenarios cover selected topics around authentication and operational security and require web application input. With the increasing number of digital identities and different identity management protocols, this is also true for identity management security. Our scenarios provide the first step. In order to teach web application security resp. identity management, additional scenarios are necessary.

\paragraph{Summary:} In summary, our three consecutive scenarios provide a first step towards detecting and mitigating brute-force attacks on authentication in web applications. It utilizes basic functionalities and tools for a better understanding and an easier start. Thereby, the participants see the effects of brute-force attacks by Wireshark and log files and simple prevention by Fail2ban. This knowledge helps to apply and evaluate tools for better discovery and prevention. This setting can be transferred to different environments by utilizing the description language. Depending on the organization and infrastructure, other tools, such as security information and event management (SIEM), and Windows OS may be available. For further training, more scenarios are required.

\section{\uppercase{Conclusions}}
\label{sec:conclusion}

The number of digital accounts is ever-increasing. The same is the case with attacks on them, ranging from brute-force attacks to sophisticated supply chain attacks targeting active directory environments. In order to train beginners, we designed a series of blue team scenarios with the topic of brute-force attacks on authentication in a web application. In the first step, we evaluated related work. With the help of the results obtained, three consecutive scenarios for the cyber range were designed and implemented. These scenarios each cover a specific part of the learning content and lay the foundation for beginners and those interested in protection against brute-force attacks. Nonetheless, more scenarios are required to train future system administrators. Finally, the open-source scenarios were assessed in a small-scale training and a discussion.

In future work, we plan to extend the scenarios to cover different attacks on identities and identity management systems, such as the more advanced attacks of Kerberoasting, Golden Ticket, and Golden SAML, and other OSs, in particular Windows. In addition, we will evaluate and improve our description language with these scenarios and discuss it with other experts. For the brute-force scenarios, regular traffic was generated with Selenium. We want to investigate other techniques and represent the traffic more realistically for a better training setup in future work.

\bibliographystyle{apalike}
{\small
\bibliography{example}}

\begin{thebibliography}{}

\bibitem[Adeleke et~al., 2022]{10.1145/3488375}
Adeleke, O.~A., Bastin, N., and Gurkan, D. (2022).
\newblock {Network Traffic Generation: A Survey and Methodology}.
\newblock {\em ACM Comput. Surv.}, 55(2).

\bibitem[{Adobe}, 2023]{xampp}
{Adobe} (2023).
\newblock {XAMPP Apache + MariaDB + PHP + Perl}.
\newblock \url{https://www.apachefriends.org/index.html}.
\newblock accessed \today.

\bibitem[Al-Mohannadi et~al., 2016]{AllgemeineAnalyse}
Al-Mohannadi, H., Mirza, Q., Namanya, A., Awan, I., Cullen, A., and Disso, J.
  (2016).
\newblock {Cyber-Attack Modeling Analysis Techniques: An Overview}.
\newblock In {\em Proceedings of the 4th International Conference on Future
  Internet of Things and Cloud Workshops (FiCloudW)}, pages 69--76. IEEE.

\bibitem[Arshad et~al., 2021]{9454094}
Arshad, S., Alam, M., Al-Kuwari, S., and Khan, M. H.~A. (2021).
\newblock {Attack Specification Language: Domain Specific Language for Dynamic
  Training in Cyber Range}.
\newblock In {\em Proceedings of the 12th Global Engineering Education
  Conference (EDUCON)}, pages 873--879. IEEE.

\bibitem[{FoxyProxy}, 2023]{foxy}
{FoxyProxy} (2023).
\newblock {FoxyProxy}.
\newblock \url{https://getfoxyproxy.org}.
\newblock accessed \today.

\bibitem[G\'{a}likov\'{a} et~al., 2021]{gamifcation}
G\'{a}likov\'{a}, M., \v{S}v\'{a}bensk\'{y}, V., and Vykopal, J. (2021).
\newblock {Toward Guidelines for Designing Cybersecurity Serious Games}.
\newblock In {\em Proceedings of the 52nd Technical Symposium on Computer
  Science Education (SIGCSE)}, page 1275. Association for Computing Machinery.

\bibitem[Kaschow et~al., 2017]{CyberRangeDefUniBw}
Kaschow, R., Hanka, O., Knüpfer, M., and Eiseler, V. (2017).
\newblock {Cyber Range: Netzverteidigung trainieren mittels Simulation}.
\newblock In {\em Proceedings of the D•A•CH Security 2017}, pages 126--137.
  syssec.

\bibitem[Kn{\"u}pfer et~al., 2020]{CyberRangeDocUniBw}
Kn{\"u}pfer, M., Bierwirth, T., Stiemert, L., Schopp, M., Seeber, S., P{\"o}hn,
  D., and Hillmann, P. (2020).
\newblock {Cyber Taxi: A Taxonomy of Interactive Cyber Training and Education
  Systems}.
\newblock In Hatzivasilis, G. and Ioannidis, S., editors, {\em Model-driven
  Simulation and Training Environments for Cybersecurity}, pages 3--21, Cham.
  Springer International Publishing.

\bibitem[Koutsouris et~al., 2021]{9527946}
Koutsouris, N., Vassilakis, C., and Kolokotronis, N. (2021).
\newblock {Cyber-Security Training Evaluation Metrics}.
\newblock In {\em Proceedings of the 1st International Conference on Cyber
  Security and Resilience (CSR)}, pages 192--197. IEEE.

\bibitem[Lopez-Araiza and Cankaya, 2017]{LopezAraiza2017}
Lopez-Araiza, C. and Cankaya, E. (2017).
\newblock {A Comprehensive Analysis of Security Tools for Network Forensics}.
\newblock {\em Journal of Medical - Clinical Research \& Reviews}, 1(3):1--9.

\bibitem[Lord, 2020]{Nate}
Lord, N. (2020).
\newblock {Uncovering Password Habits: Are Users’ Password Security Habits
  Improving?}
\newblock
  \url{https://digitalguardian.com/blog/uncovering-password-habits-are-users-password-security-habits-improving-infographic}.
\newblock accessed \today.

\bibitem[{Masaryk University}, 2022]{kypo}
{Masaryk University} (2022).
\newblock {KYPO Cyber Range Platform}.
\newblock \url{https://docs.crp.kypo.muni.cz}.
\newblock accessed \today.

\bibitem[Meyer, 2021]{logfilelitrech}
Meyer, R. (2021).
\newblock {Detecting Attacks on Web Applications from Log Files}.
\newblock techreport, SANS Institute.

\bibitem[Mohammed et~al., 2017]{bruteforceanawithws}
Mohammed, M.~A., Degadzor, A.~F., Effrim, B.~F., and Appiah, K.~A. (2017).
\newblock {Brute Force Attack detection and prevention on a network using
  wireshark analysis}.
\newblock {\em International Journal of Engineering Sciences \& Research
  Technology}, 6(6):26--37.

\bibitem[Nagarajan et~al., 2012]{6392562}
Nagarajan, A., Allbeck, J.~M., Sood, A., and Janssen, T.~L. (2012).
\newblock Exploring game design for cybersecurity training.
\newblock In {\em Proceedings of the International Conference on Cyber
  Technology in Automation, Control, and Intelligent Systems (CYBER)}, pages
  256--262. IEEE.

\bibitem[{Offensive Security}, 2023a]{tools}
{Offensive Security} (2023a).
\newblock {All Kali Tools}.
\newblock \url{https://www.kali.org/tools/all-tools/}.
\newblock accessed \today.

\bibitem[{Offensive Security}, 2023b]{pen}
{Offensive Security} (2023b).
\newblock {PEN-103 Modules}.
\newblock
  \url{https://portal.offensive-security.com/courses/pen-103/books-and-videos/modules}.
\newblock accessed \today.

\bibitem[{Port Swigger}, 2022]{BurpSuiteInfos}
{Port Swigger} (2022).
\newblock {Burp Suite documentation}.
\newblock \url{https://portswigger.net/burp/documentation}.
\newblock accessed \today.

\bibitem[Ramya et~al., 2017]{seleniuminf}
Ramya, P., Sindhura, V., and Sagar, P.~V. (2017).
\newblock Testing using selenium web driver.
\newblock In {\em Proceedings of the 2nd International Conference on
  Electrical, Computer and Communication Technologies (ICECCT)}, pages 1--7.
  IEEE.

\bibitem[Shin et~al., 2019]{8992693}
Shin, S., Seto, Y., Kasai, Y., Ka, R., Kuroki, D., Toyoda, S., Hasegawa, K.,
  and Midorikawa, K. (2019).
\newblock {Development of Training System and Practice Contents for
  Cybersecurity Education}.
\newblock In {\em Proceedings of the 8th International Congress on Advanced
  Applied Informatics (IIAI-AAI)}, pages 172--177. IEEE.

\bibitem[Simic, 2019]{apacheaccerr}
Simic, S. (2019).
\newblock {How to View Apache Access \& Error Logs}.
\newblock \url{https://phoenixnap.com/kb/apache-access-log}.
\newblock accessed \today.

\bibitem[Srinivasa~Rao and Pais, 2017]{10.1145/3055186.3055188}
Srinivasa~Rao, R. and Pais, A.~R. (2017).
\newblock {Detecting Phishing Websites Using Automation of Human Behavior}.
\newblock In {\em Proceedings of the 3rd Workshop on Cyber-Physical System
  Security (CPSS)}, page 33–42. Association for Computing Machinery.

\bibitem[Stu et~al., 2022]{steiner}
Stu, S., Ananth, J., and de~Leon~Daniel, C. (2022).
\newblock {A Survey of Cloud-hosted, Publicly-available, Cyber-ranges for
  Educational Institutions}.
\newblock {\em Journal of Computing Sciences in Colleges}, 38.

\bibitem[Subaşu et~al., 2017]{8166396}
Subaşu, G., Roşu, L., and Bădoi, I. (2017).
\newblock {Modeling And Simulation Architecture For Training In Cyber Defence
  Education}.
\newblock In {\em Proceedings of the 9th International Conference on
  Electronics, Computers and Artificial Intelligence (ECAI)}, pages 1--4. IEEE.

\bibitem[Tanaka et~al., 2020]{9155736}
Tanaka, T., Niibori, H., Shiyingxue, L., Nomura, S., Nakao, T., and Tsuda, K.
  (2020).
\newblock {Selenium based Testing Systems for Analytical Data Generation of
  Website User Behavior}.
\newblock In {\em Proceedings of the 13th International Conference on Software
  Testing, Verification and Validation Workshops (ICSTW)}, pages 216--221.
  IEEE.

\bibitem[{tryhackme}, 2021]{thmintruder}
{tryhackme} (2021).
\newblock {Burp Suite: Intruder}.
\newblock \url{https://tryhackme.com/room/burpsuiteintruder}.
\newblock accessed \today.

\bibitem[Usha et~al., 2010]{wiresharkbegr}
Usha, B., Ashutosh, V., and Saxena, M. (2010).
\newblock {Evaluation of the Capabilities of WireShark as a tool for Intrusion
  Detection}.
\newblock {\em International Journal of Computer Applications}, 6(7):1--5.

\bibitem[Vykopal et~al., 2017]{icsoft17}
Vykopal, J., Oslejsek, R., Celeda, P., Vizvary, M., and Tovarnak, D. (2017).
\newblock {KYPO Cyber Range: Design and Use Cases}.
\newblock In {\em Proceedings of the 12th International Conference on Software
  Technologies (ICSOFT)}, pages 310--321. INSTICC, SciTePress.

\bibitem[Walden, 2008]{10.1145/1414558.1414607}
Walden, J. (2008).
\newblock {Integrating Web Application Security into the IT Curriculum}.
\newblock In {\em Proceedings of the 9th SIGITE Conference on Information
  Technology Education (SIGITE)}, page 187–192, New York, NY, USA.

\end{thebibliography}

\end{document}